\documentclass{aastex}
\usepackage{spr-astr-addons}
\usepackage{url}
\usepackage{graphicx,times}

\RequirePackage{color}

\begin{document}

\title{The influence of outflow and global magnetic field on the structure and spectrum of resistive CDAFs}
\shorttitle{Short article title}

\author{Maryam Ghasemnezhad\altaffilmark{1}} \and \author{Shahram Abbassi\altaffilmark{1,2}}
\email{m.ghasemnezhad@uk.ac.ir}
\affil{abbassi@um.ac.ir}

\altaffiltext{1}{Faculty of physics, Shahid Bahonar University of Kerman, Kerman, Iran}
\altaffiltext{2}{Department of Physics, School of Sciences, Ferdowsi University of Mashhad, Mashhad, 91775-1436, Iran}
\email{abbassi@um.ac.ir}

\begin{abstract}
    We examine the effects of a global magnetic field and outflow on radiatively inefficient accretion flow (RIAF) in the presence of magnetic resistivity.  We find a self-similar solutions  for the height integrated equations that govern the behavior of the flow. We use the mixing length mechanism for studying the convection parameter. We adopt a radius dependent mass accretion rate as $\dot{M}=\dot{M}_{out}{(\frac{r}{r_{out}})^{s}}$ with $s> 0$ to investigate the influence of outflow on the structure of inflow where $s$ is a constant and indication the effect of wind. Also, we have studied the radiation spectrum and temperature of CDAFs. The thermal bermsstrahlung emission as a radiation mechanism is taken into account for calculating the spectra emitted by the CDAFs. The energy that powers bremsstrahlung emission at large radii is provided by convective transport from small radii and viscous and resistivity dissipation. Our results indicate that the disc rotates slower and accretes faster, it becomes hotter and thicker for stronger wind. By increasing all component of magnetic field, the disc rotates faster and accretes slower while it becomes hotter and thicker. We show that the outflow parameter and all component of magnetic field have the same effects on the luminosity of the disc. We compare the dynamical structure of the disc in two different solutions (with and without resistivity parameter). We show that only the radial infall velocity and the surface density could changed by resistivity parameter obviously. Increasing the effect of wind increases the disc's temperature and  luminosity of the disc. The effect of magnetic field is similar to the effect of wind in the disc's temperature and luminosity of the disc, but the influence of resistivity on the observational properties
is not evident. 
 
\end{abstract}



\section{INTRODUCTION}
 Black hole accretion discs provide the most powerful energy production mechanism in the universe. It is well accepted that many astrophysical objects are powered by black hole accretion. The standard geometrically thin, optically thick accretion disc model can be successfully explain many observational features of X-ray binaries, but it is unable to explain observations of low-luminosity X-ray binaries and AGNs accretion discs. A particular example of such low luminous sources is our galactic center, Sagittarius, with host a $2\times 10^6$ solar mass black hole with luminosity well below the estimated value based on standard model (Melia \& Falcke 2001). At low luminosities (less than a few percent of the Edington luminosity), black holes can accrete via advection dominated accretion flows (ADAFs) (Ichimaru 1977, Narayan \& Yi 1994, Kato, Fukue \& Mineshige 2008 and Yuan \& Narayan 2014 for review). In such a flow, radiative losses are small compare to viscously heating because of low particle density of accreting flow at low accretion rate. Consequently, most of the energy released via viscosity is stored as entropy and transport inward with accretion. ADAFs are optically thin, geometrically thick and hot (compare the virial temperature of the gas in the flow) and radiate mostly in X-ray band (see Narayan et al. 1996). In the past decades the ADAFs models have captured great attentions and rapid progress has been made. 

At the same time as ADAFs model was introduced, it was realized that they are likely to be unstable against convection in the radial direction. Because of low radiative efficiency in hot accretion flow, since the gas is heated but hardly cools, the entropy increases with decreasing radius. Hot accretion flows are therefore potentially unstable to convection. However, according some numerical simulations (e.g. Stone et al. 1999; Narayan et al. 2000) there are some debates about whether convection exists in hot accretion flow or not. Some of them have clearly shown that convection in an MHD accretion flow likely does not exist (Pen et al. 2003; Narayan et al. 2012; Yuan et al. 2012b). But some other series of numerical simulations reveal that the convection instability likely occurs in hot accretion flows (Igumenshchev, Chen \& Abramowicz 1996, Igumenshchev \& Abramowicz 1999, 2000, Stone, Pringle \& Begelman 1999, Yuan \& Bu 2010). Of course, some uncertainties still exist and we can't conclude the non-existence of convection (see discussion in Yuan \& Narayan 2014), thus it is still feasible to study convection and it worth to study CDAFs. Narayan, Igumenshchev \& Abramowicz (2000) and Quataert \& Gruzinuv (2000) introduced analytical model based on self-similar solution which was called convection dominated accretion flows (CDAFs).  In particular,  Igumenshchev, \& Abramowicz (1999, 2000) have been point out that the ADAFs becomes convectively unstable whenever the viscous parameter $\alpha \le 0.1$. On the other hand, Narayan et. al (2000), Quataret \& Gruzinuv (2000) based on self-similar solutions have shown that CDAF consist of a hot plasma about virial temperature and have a flattened time-averaged radial density profile, $\rho\propto r^{-\frac{1}{2}}$, where much flatter than usual ADAFs with $\rho\propto r^{-\frac{3}{2}}$. In CDAFs the most part of the energy which realized in inner most region of accretion flow is transport outward by convection. 

Mass loss mechanism (in the form of wind or outflow) is an interesting phenomenon in the structure and evolution of the accretion discs. The existence of wind and outflow has been observationally verified in various astronomical objects like AGNs and YSOs (Whelan et al. 2005, Bally et al. 2007). \textbf{On the other hand several numerical simulation have been performed and they clearly confirm the existence of outflow in such systems (Yuan et al. (2012a, 2012b), Narayan et al. (2012), Li, Ostriker \& Sunyaev (2013), Yuan et al. (2015), Bu et al. (2016a, 2016b).} In these objects some part of angular momentum of the accretion flow will dissipated outward in the form of wind and jet. For generating outflow various driving forces are proposed, such as thermal, radiative and magnetic field. The wind mechanism has been investigated by many others (Meier 1979, Fukue 1989, Abbassi et al. 2008, 2010, Ghasemnezhad \& Abbassi 2016). The effect of magnetic field on the disc were also studied ( see Balbus \& Hawley 1998, Kaburaki 2000, Shadmehri \& Khajenabi 2005, Abbassi et al. 2008, Ghasemnezhad et al. 2012, 2013,  Samadi et al. 2014, 2016, Bu et al. 2009, Soria et al. 1997). The effect of large scale magnetic field on the physical properties of CDAFs with hydrodynamically driven wind have been investigated by Abbassi \& Mosallanezhad (2012, here after AM12).

The magnetic field have several effect in the dynamical and observational appearance of the discs such as:  the formation of wind/jet, the interaction of discs and black holes and synchrotron emission. The traditional view of the magnetic field in the accretion disc is that the magnetic field is not completely frozen into the  accreting matter. The fluid is not a perfect conductor, so the magnetic field advected inward by accretion and diffused by viscosity and resistivity (Guan \& Gammie 2009). The resistivity diffusion of magnetic field is important in accretion disc and the simulations of local shearing box have indicated that the resistive dissipation increases the linear growth rate of magneto rotational instability (MRI) (Fleming et al. 2000). It will be interesting to study the effect of resistivity on optically thin ADAFs with convection, outflow and global magnetic field. Faghei \& Omidvand (2012, hear after FO12) studied radial self similar solution of accretion flow in the presence of toroidal magnetic field, convection and resistivity. They ignored the effect of outflow and global magnetic field. AM12 studied the self similar solution of CDAFs with a global magnetic field and outflow. We have improved AM12 paper by adding the magnetic resistivity parameter and then have compared two solutions.

The main aim of our present work is highlighting observational consequences of CDAF models, focusing in particular on power spectra. Our results are similar to those of Blandford \& Begelman(1999), AM12 who have assumed that a significant fraction of mass in an ADAF would be lost to outflow/wind, rather than accreting onto central object.  In ADAFs the importance of outflows can be shown by a radial density profile as  ($\rho \propto r^{s-\frac{3}{2}}$) which $(0<s<1)$. The density profile in CDAFs is equivalent to $s=1$ ($\rho \propto r^{-\frac{1}{2}}$). In order to capture many feature of CDAFs we have considered various values of $s$  in this study. 

In CDAFs the convection motions transport a luminosity $L_{c}\approx (10^{-3}-10^{-2})\dot{M} c^{2}$ from small to large radii. The most of the energy that transport outward by convection can be radiated from the outer regions of the flow as thermal bremsstrahlung emission which is a function of temperature and density of accreting gas (Igumenshchev, \& Abramowicz (2000) , Ball et al. 2001). We assumed the same mechanism in our study.

This paper organized as follows. The basic equations and assumptions are presented in section 2. Self-similar solutions are presented in section 3. The radiation properties of CDAFs are discussed in section 4. We show the result in section 5 and finally we present the
summary and conclusion in section 6.

\section{The Basic Equations}
We use the cylindrical coordinates $(r, \varphi, z)$ to write the MHD equations of steady state and axi-symmetric ($\frac{\partial}{\partial
\phi}=\frac{\partial}{\partial t}=0$) hot accretion flow around compact black hole of mass
$M_{\star}$. Following AM12, we assume a magnetic field with three components $(B_{r}, B_{\varphi},
B_{z})$. We have vertically integrated the equations and then all our physical variables become only a function of radial distances, $r$. Moreover, the disc suppose to have Newtonian gravity in radial direction and also we neglect the self-gravity of the discs and the general relativistic effects. The disc is supposed to turbulent and possesses an effective turbulent viscosity. We adopt $\alpha$ -prescription for viscosity of rotating gas in accretion flow. The convection, outflow and magnetic field and its correspond resistivity are important to transfer of energy and angular momentum in disc.

The equation of continuity gives:

\begin{equation}
\frac{\partial}{\partial r}(r \Sigma V_r )+\frac{1}{2\pi}\frac{\partial \dot{M}_\mathrm{w}}{\partial r}=0
\end{equation}
where $V_{r}$ is the accretion velocity ($V_{r}<0$) and
$\Sigma=2\rho H$ is the surface density at a cylindrical radius
$r$. $H$ is the disc half-thickness and $\rho$ is the density.
Mass-loss rate by wind/outflow is represented by $\dot{M}_\mathrm{w}$. So

\begin{equation}
\dot{M}_\mathrm{w}=\int 4\pi r'\dot{m}_\mathrm{w}(r')dr',
\end{equation}
where $\dot{m}_\mathrm{w}(r)$ is the mass-loss per unit area
from each disc face. Similar to Blandford \& Begelman (1999) and AM12, we write the dependence of accretion rate  as follows,

\begin{equation}
\dot{M}=-2\pi r \Sigma V_r =\dot{M}_{out}(\frac{r}{r_{out}})^{s}
\end{equation}

where $\dot{M}_{out}$ is the mass accretion rate at the outer adge of the disc ($r_{out}$) (Blandford \& Begelman 1999) and $s$ is a constant with order of unity. Considering equation (1-3), we can write 
\begin{equation}
\dot{m}_\mathrm{w}=\frac{s \dot{M}_{out}}{4 \pi r_{out}^{2}}(\frac{r}{r_{out}})^{s-2}
\end{equation}
The equation of motion in the radial direction is:

\begin{displaymath}
V_r \frac{\partial V_r}{\partial
r}=\frac{V_{\varphi}^{2}}{r}-\frac{GM_{\star}}{r^2}-\frac{1}{\Sigma}\frac{d}{dr}(\Sigma
c_\mathrm{s}^{2})
\end{displaymath}

\begin{equation}
-\frac{c_\mathrm{\varphi}^{2}}{r}-\frac{1}{2\Sigma}\frac{d}{dr}(\Sigma c_\mathrm{\varphi}^{2}+\Sigma c_\mathrm{z}^{2})
\end{equation}

 where $V_{\varphi}$, $G$ and $c_\mathrm{s}$ are the rotational velocity of
 the flow, the gravitational constant and sound speed respectively. The sound speed is
 defined as 
 $c_\mathrm{s}^{2}=\frac{p_\mathrm{gas}}{\rho}$
 where $p_\mathrm{gas}$ is the gas pressure. Following AM12 and Zhang \& Dai (2008), we introduce three component of Alfven sound speed            $c_\mathrm{r},c_\mathrm{\varphi}$ and $c_\mathrm{z}$ as:
 
 \begin{equation}
 c_\mathrm{r,\varphi,z}^{2}=\frac{B_{r,\varphi,z}^{2}}{\textbf{4}\pi\rho}=\frac{2p_\mathrm{mag_{r,\varphi,z}}}{\rho}
 \end{equation}
 
 where $B_{r,\varphi,z}$ and $p_\mathrm{mag_{r,\varphi,z}}$ are three components of magnetic field and magnetic pressure
 respectively.

 By integration over $z$ of the azimuthal equation of motion gives.
 
 \begin{displaymath}
   \Sigma V_r \frac{d}{dr}(rV_\varphi)=-\frac{1}{r}\frac{d}{dr}(J_{vis})-\frac{1}{r}\frac{d}{dr}(J_{con})-  \frac{\Omega(lr)^{2}}{2\pi}\frac{d\dot{M_{w}}}{dr}
 \end{displaymath}
 
\begin{equation}
   +r \sqrt{\Sigma} c_\mathrm{r}\frac{d}{dr}(\sqrt{\Sigma} c_\mathrm{\varphi})+\Sigma c_\mathrm{r} c_\mathrm{\varphi} 
\end{equation}

 where $\Omega(=\frac{V_\varphi}{r})$ and are the angular and Keplerian velocities respectively. The third term on the
right hand side shows the angular momentum carried a way by
wind/outflow materials. Knigge (1999) define the $l$ parameter as the length of the
rotational lever-arm that allows we have several types of accretion
disc winds models. The parameter $l=0$ corresponds to a
non-rotating wind and the angular momentum is not extracted by
the wind and the disc losses only mass because of the wind while
$l=1$ represents outflowing materials that carries away the specific angular momentum $(r^{2}\Omega)$. This latter would be most fitting value for radiation-driven wind (Proga et al. 1998). Centrifugally driven MHD wind/outflow are correspond to $l>1$ and it would be able to remove a lot of angular momentum of the discs.

 The $J_{vis}$ and $J_{con}$ are the viscous and convective angular momentum fluxes respectively that define as follow:
\begin{equation}
J_{vis}=-r^{3}\nu
\Sigma\frac{d\Omega}{dr}
\end{equation}
and 
\begin{equation}
J_{con}=-\nu_{con} 
\Sigma r^{3\frac{(1+g)}{2}}\frac{d}{dr}(\Omega  r^{3\frac{(1-g)}{2}})
\end{equation}

where $\nu$ is the kinematic viscosity coefficient and formalized by 
Shakura \& Sunyaev (1973) as:
\begin{equation}
 \nu=\alpha c_{s}H
 \end{equation}
where $\alpha$ is a constant less than unity that has called the viscous parameter. Also we have formalized all of the turbulence in our system like convective diffusion and resistivity similar to viscosity turbulence. So,
\begin{equation}
 \nu_{con}=\alpha_{c} c_{s}H
 \end{equation}
 where $\alpha_{c}$ is the dimensionless convective parameter and we get it according the mixing length theory and $g$ is an index for determining the condition for angular momentum transportations. There are several possibilities for transporting of angular momentum by convection (Narayan et al. 2000) which is depends on the magnitude of $g$ parameter. Generally, convection transports angular momentum inward or outward for $g < 0 $ or $g > 0$, respectively while $g = 0$ corresponds to zero angular momentum
transportation (Narayan et al. 2000). When $g=1$ the convection behaves like turbulence viscosity but if $g=-\frac{1}{3}$ the convection transport angular momentum inward. In this paper we consider the convective angular momentum flux as:
\begin{equation}
J_{con}=-r\nu_{con}
\Sigma\frac{d}{dr}(r^{2}\Omega)
\end{equation}
where is correspond to $g=-\frac{1}{3}$ and represents that the convective angular momentum flux is oriented down the specific angular momentum gradient. It means that convection tries to drive the system toward a stat of uniform specific angular momentum and consequently it corresponds to an inward angular momentum transportation (Narayan et al. 2000). We have determined the convective turbulence parameter $\alpha_{c}$ be the mixing length approximation.  It can be imagined that a convective differentially-rotating fluid include of many independent fluid blobs. following Grossman et al. (1993) the convectively turbulence viscosity is defined as 
\begin{equation}
 \nu_{turb}=\sigma L_{M}  
 \end{equation}
 where $\sigma$ is the velocity dispersion of the blobs and $L_{M}$ is the characteristic mixing length corresponding to effective mean free path of the blobs. So we can write the $\nu_{con}$ as follows (Lu et al. 2004):
 \begin{equation}
 \nu_{con}=\frac{L_{M}^{2} }{4} (-N^{2}_{eff})^{\frac{1}{2}}
 \end{equation}
 Here $N_{eff}$ is the effective frequency of the convective blobs and it will be:
 
 \begin{equation}
 N^{2}_{eff}=N^{2}+k^{2}
 \end{equation}
 where $N$ and $k$ are the Brunt-Vaisala frequency and epicyclic frequency respectively, which are defined
as
 \begin{equation}
 N^{2}=-\frac{1}{\rho}\frac{ dp_{g}}{dr}\frac{d}{dr}\ln(\frac{p^{\frac{1}{\gamma}}_{g}}{\rho})
 \end{equation}
 and 
  \begin{equation}
 k^{2}=2\Omega ^{2}\frac{d \ln(r^{2}\Omega)}{d lnr}
 \end{equation}
 Also the characteristic mixing length $L_{M}$ could be written in terms of the pressure scale height $(H_{p}=-\frac{d r}{d \ln p})$ and the dimensionless mixing length parameter $l_{m}$ as bellow:
 \begin{equation}
 L_{M}=2^{-\frac{1}{4}}l_{M}H_{p}
 \end{equation}
 
  We have adopt $l_{M}=\sqrt{2}$ as it was estimated by Narayan et al. (2000) and Lu et al. (2004). Convection is present whenever $N^{2}_{eff}< 0$. $\alpha_{c}$ can be written in the form similar to normal viscosity as:
 \begin{equation}
 \alpha_{c}=\frac{\nu_{con}}{c_{s} H}
 \end{equation}
 
By integrating along $z$ of the hydrostatic balance, we have:
\begin{equation}
\Omega^{2}_{k} H^{2}-\frac{c_\mathrm{r}}{\sqrt{\Sigma}}\frac{d}{dr}(\sqrt{\Sigma} c_\mathrm{z})H = c^{2}_\mathrm{s}+\frac{1}{2}(c^{2}_\mathrm{r}+c^{2}_\mathrm{\varphi})
\end{equation} 

In order to complete the problem we need to introduce energy
equation. We assume the generated energy due to viscosity and resistivity into the volume is balanced by the advection cooling, outward energy by convection
and energy loss of outflow $(Q^{adv}+Q^{rad}+Q^{conv}+Q^{wind}=Q^{diss})$. Thus,
\begin{displaymath}
\Sigma V_{r} T \frac{d S}{dr}+\frac{1}{r}\frac{d}{dr}(r F_{con})= f(\nu+g\nu_{con})\Sigma r^{2} (\frac{d \Omega}{dr})^{2}+\frac{\eta}{4\pi}\mathrm{J}^{2}
\end{displaymath}
\begin{equation}
-\frac{1}{2}\zeta \dot{m}_\mathrm{w}(r)V_\mathrm{k}^{2}(r)
\end{equation}

where $F_{con}$, $S$, $f$ and $T$ are the convective energy flux, the specific entropy, advection parameter and temperature respectively. Also we consider $Q^{diss}-Q^{rad}=fQ^{diss}$. Their corresponding relations are:
\begin{equation}
F_{con}=-\nu_{con} \Sigma T \frac{d S}{dr}
\end{equation}
where 
\begin{equation}
T\frac{d S}{dr}=\frac{1}{\gamma-1}\frac{d c^{2}_{s}}{dr}-\frac{c^{2}_{s}}{\rho}\frac{d \rho}{dr}
\end{equation}
here $\gamma$ is the specific energy heats, $J=\nabla \times B $ is the current density and $\eta$ is the magnetic diffusivity due to turbulence. The two first term on the right hand side of the energy equation corresponds to the dissipation energy by viscosity, convection and resistivity $Q_{diss}= f(\nu+g\nu_{con})\Sigma r^{2} (\frac{d \Omega}{dr})^{2}+\frac{\eta}{4\pi}\mathrm{J}^{2}$. We can write the magnetic resistivity turbulence in the form of viscosity and convection turbulence as we was stated as:
\begin{equation}
\nu=P_{m}\eta=\alpha c_{s} H
\end{equation}
where $P_{m}$ is the magnetic Prandt number of the turbulence, which adopted to be a constant less than unity, $\eta$ is the magnetic diffusivity (Shadmehri 2004). 
The last term on the right hand side
of the energy equation represents the energy loss due to wind or
outflow (Knigge 1999). In our model $\zeta$ is a free and
dimensionless parameter. The large $\zeta$ corresponds to more
energy extraction from the disc because of wind (Knigge 1999).

Finally since we consider three components of magnetic
field, the three components of induction equation can be written as:
\begin{equation}
\dot{B}_{r}=0
\end{equation}

\begin{equation}
\dot{B}_{\varphi}=\frac{d}{dr}[V_\varphi B_r - V_r B_\varphi+\frac{\eta}{r}\frac{d}{dr}(r B_\varphi)]
\end{equation}

\begin{equation}
\dot{B}_{z}=-\frac{d}{dr}[rV_r B_z -\eta r\frac{d B_z}{dr}]
\end{equation}
where $\dot{B}_ r,\varphi,z$ is the field escaping/creating rate due
to magnetic instability or dynamo effect. Now we have a set of MHD equations that control the structure of magnetized CDAFs. The solutions of these coupled equation are strongly correlated to given values of viscosity, connectivity, magnetic field strength, $\beta_{r, \phi, z}$ and degree of advection $f$. In the next section we will demonstrate the self-similar solution of this MHD equations.

\section{Self-Similar Solutions}
The basic equations of our models was discussed in the last section. We use the self similar method to solving above complicated differential equations. This powerful technique is a dimensional analysis and scaling law and is widely used in astrophysical fluid mechanics. Following to AM12, and other similar works (Ghanbari et al 2009 and samadi et al 2014, 2016), self-similarity in the radial direction is assumed:
\begin{equation}
\Sigma=c_{0}\Sigma_{out} (\frac{r}{r_{out}})^{s-\frac{1}{2}}
\end{equation}
\begin{equation}
V_r(r)=-c_1 \sqrt{\frac{G M_{\ast}}{r_{out}}}(\frac{r}{r_{out}})^{-\frac{1}{2}}
\end{equation}
\begin{equation}
V_\varphi(r)= r \Omega (r) =c_2 \sqrt{\frac{G M_{\ast}}{r_{out}}}(\frac{r}{r_{out}})^{-\frac{1}{2}}
\end{equation}
\begin{equation}
c_\mathrm{s}^{2}=c_3 \frac{G M_{\ast}}{r_{out}}(\frac{r}{r_{out}})^{-1}
\end{equation}
\begin{equation}
c_\mathrm{r, \varphi, z}^{2} =\frac{B^{2}_{r, \varphi, z}}{4\pi\rho}= 2 \beta_{r, \varphi, z} c_3 \frac{G M_{\ast}}{r_{out}}(\frac{r}{r_{out}})^{-1}
\end{equation}
\begin{equation}
H(r)=c_{4}r_{out}(\frac{r}{r_{out}})
\end{equation}
\begin{equation}
\rho=\frac{1}{2}\frac{c_0 \Sigma_{out}}{c_4}\frac{1}{r_{out}}(\frac{r}{r_{out}})^{s-\frac{3}{2}}
\end{equation}
\begin{equation}
B_{r,\varphi,z}= 2\sqrt{\frac{\pi\beta_{r, \varphi, z} c_3 c_0 \Sigma_{out} G M_\ast}{c_4}} \frac{1}{r_{out}}(\frac{r}{r_{out}})^{\frac{s}{2}-\frac{5}{4}}  
  \end{equation}
where the constant $c_0$, $c_1$, $c_2$, $c_3$ and $c_4$ are dimensionless constants and will be determined later. $\Sigma_{out}$ and $r_{out}$ have been exploited in order to write equations in non-dimensional form. Substituting the above self-similar transformation in the MHD equations of the system,
we'll obtain the following system of coupled ordinary equations, which should be solve to having
$c_0$, $c_1$, $c_2$, $c_3$ and $c_4$:
\begin{equation}
\dot{m}=c_0 c_1
\end{equation}

\begin{equation}
-\frac{1}{2}c_1^{2}=c_2^{2}-1-[(s-\frac{3}{2})+\beta_\varphi(s+\frac{1}{2})+(s-\frac{3}{2})\beta_z]c_3
\end{equation}

 \begin{equation}
 -\frac{1}{2}c_1 c_2=-\frac{3}{2}(s+\frac{1}{2})(\alpha+g\alpha_c)c_2 c_4 \sqrt{c_3}+(s+\frac{1}{2})c_3\sqrt{B_r B_\varphi}-s l^2 c_1 c_2
 \end{equation}
 \begin{equation}
 c_4=\frac{1}{2}(s-\frac{3}{2})\sqrt{B_r B_z}+\frac{1}{2}\sqrt{(s-\frac{3}{2})^{2} c^{2}_{3} B_r B_z+4(1+B_r+B_\varphi)c_3}
 \end{equation}
 \begin{displaymath}
 (\frac{1}{\gamma-1}+s-\frac{3}{2})[(s-1)\alpha_c c^{\frac{3}{2}}_{3}c_4+c_1 c_3]=-\frac{1}{4}s \zeta c_1
 \end{displaymath}
 \begin{displaymath}
 +f[\frac{9}{4}(\alpha+g\alpha_c)c_4 c^{2}_{2} c^{\frac{1}{2}}_{3}  + \frac{1}{2}\frac{\alpha}{P_m}c_4 c^{\frac{3}{2}}_{3}\beta_\varphi (s-\frac{1}{2})^{2}
 \end{displaymath}
 \begin{equation}
+\frac{1}{2}\frac{\alpha}{P_m}c_4 c^{\frac{3}{2}}_{3}\beta_z (s-\frac{5}{2})^{2}-\frac{\alpha}{P_m}c_4 c^{\frac{3}{2}}_{3}\sqrt{\beta_z \beta_\varphi} (s-\frac{5}{2})(s-\frac{1}{2})]
 \end{equation}
 \begin{equation}
 \alpha_c=\frac{l^{2}_M}{4\sqrt{2 c_3}c_4 (s-\frac{5}{2})^{2}}\sqrt{(s-\frac{5}{2})c_3[(s-\frac{5}{2})\frac{1}{\gamma}-(s-\frac{3}{2})]-c^{2}_{2}}
 \end{equation}
 where $\dot{m}$ is the dimensionless mass accretion rate and define as:
\begin{equation}
\dot{m}=\frac{\dot{M}_\mathrm{out}}{\pi \Sigma_{out} r_{out}\sqrt{\frac{GM_{\ast}}{r_{out}}}}
\end{equation}
Also, the field scaping/creating rate $\dot{B}_r,\varphi,z$ is written as follows:
\begin{equation}
\dot{B}_{r,\varphi,z}=\dot{B}_{0 r,0\varphi,z}(\frac{r}{r_{out}})^{\frac{s}{2}-\frac{11}{4}}
\end{equation}
 By using of the self similarity solutions, we will have:
 \begin{equation}
 \dot{B}_{r}=0
 \end{equation}
 \begin{displaymath}
 \dot{B}_{0\varphi}=\frac{1}{2}(s-\frac{7}{2})\frac{GM_{\ast}}{r^{\frac{5}{2}}_{out}}\sqrt{\frac{4\pi c_0 c_3 \Sigma_{out}}{c_4}}[c_2\sqrt{B_r}+c_1\sqrt{B_\varphi}
 \end{displaymath}
 \begin{equation}
 -(s-\frac{1}{2})\frac{\alpha c_4 \sqrt{c_3}}{2 P_m}]
 \end{equation}

 \begin{equation}
 \dot{B}_{0z}=\frac{1}{2}(s-\frac{3}{2})\frac{GM_{\ast}}{r^{\frac{5}{2}}_{out}}\sqrt{\frac{4\pi c_0 c_3 \Sigma_{out}}{c_4}}[c_1
 -(s-\frac{5}{2})\frac{\alpha c_4 \sqrt{c_3}}{2 P_m}]
 \end{equation}
 We can solve these equations numerically. The equations reduce to the equations of AM12 without the resistivity parameter $\eta=0$ or ($P_m=\infty$), . Also our equations reduce to the result of FO12 without outflow/wind parameter, radial and vertical magnetic filed.

 \section{The radiation properties of CDAFs}
 Using self-similar solutions obtained in the pervious section we will able to produce observational appearance of CDAFs. The inner part of accretion discs, where ADAFs or CDAFs conditions is valid, has a very high temperature and is moreover optically thin and magnetized. The relevant radiation processes are synchrotron emission, bremesstrahlung and modified Comptonization. Bremsstrahlung emission is the main cooling process for high temperature $T (>10^{7} K)$ plasma. In this paper, we have supposed that the bremsstrahlung radiation is the only contributor to our spectrum model.
\\
As we have shown in the last section, the density of gas in our model is :
\begin{equation}
\rho=\frac{c_0}{c_4}(\frac{1}{2}\frac{\Sigma_{out}}{r_{out}})(\frac{r}{r_{out}})^{s-\frac{3}{2}}=\frac{c_0}{c_4}\rho_{out} (\frac{r}{r_{out}})^{s-\frac{3}{2}}
\end{equation}

Following $Ball$ et al. (2001), we employ the Schawrzchild units for the radius, i.e., $R=\frac{r}{R_s}$ and $R_{out}=\frac{r_{out}}{R_s}$.  $R_s=\frac{2GM_{\ast}}{c^{2}}=2.95 \times 10^{5} m \  cm$ is the Schawrzchild radius and $c$ and $m$ are light speed and the black hole mass in solar units $(=\frac{M_{\ast}}{M_{\odot}})$ respectively. So we can write the density as:
\begin{equation}
\rho=\frac{c_0}{c_4}\rho_{out} R^{\frac{3}{2}-s}_{out} (R)^{s-\frac{3}{2}}=\frac{c_0}{c_4}\rho_0 (R)^{s-\frac{3}{2}}
\end{equation}
where $\rho_{0}=\rho_{out} R^{\frac{3}{2}-s}_{out}$.
As we stated in introduction, for $s=1$ the ADAF(+wind+convection) solutions include of CDAF models.

 In this model, we can estimate the temperature of the dics as (Akizuki \& Fukue
2006):
\begin{equation}
\frac{\Re}{\bar{\mu}}T=c_\mathrm{s}^{2}=c_3\frac{GM_{\star}}{r}
\end{equation}
 where $\Re$ the gas constant and
 $\bar{\mu}$ the mean molecular weight
($\bar{\mu}=0.5$). So,
 \begin{displaymath}
 T=c_3\frac{c^{2}\bar{\mu}}{2 \Re}(\frac{r}{R_\mathrm{s}})^{-1}=2.706 \times10^{12}c_3(\frac{r}{R_\mathrm{s}})^{-1}
 \end{displaymath}
 \begin{equation}
=2.706 \times10^{12}c_3(R)^{-1}=T_0 c_3 (R)^{-1}
\end{equation}
 In this formula the coefficient $c_3$ implicitly
depends on the wind, magnetic diffusion, magnetic field,
advection and viscosity parameters, ($s, \eta (or P_m), \beta_{r,\varphi,z}, f, \alpha$).
 	Total emissivity due the bremsstrahlung emission is (Rybicki \& Lightman 1986):
 	\begin{displaymath}
 		q_{bremss}(T,\nu)=2.4\times 10^{10} T^{-\frac{1}{2}}\rho^{2}\exp(\frac{-h\nu}{KT}) G_b
 	\end{displaymath}
 	\begin{equation}
 (erg s^{-1} cm^{-3}Hz^{-1}) 
 	\end{equation}
 	where $G_b$, $K$ and $h$ are the Gaunt factor (and is around $1$), the Boltzmann constant and Planck's constant respectively. We have ignored a weak frequency-dependent Gaunt
factor.
 	As we see the density $\rho$ instead of temperature is the dominant factor in bremsstrahlung emission. As the above equation, the bremsstrahlung emission in CDAFs is a dominant process  because of the less steep density profile.  Also the spectral structure of CDAFs are similar to the ADAFs with outflow solutions (Ball et al. 2001). Therefore we consider bremsstrahlung X-ray emission from the outer parts of CDAFs.
 	 By integrating over the whole frequency, we have:
 	 \begin{equation}
 	q_{bremss}(T)=5 \times 10^{20} T^{\frac{1}{2}}\rho^{2} G_b(erg s^{-1} cm^{-3})
 	\end{equation}
 	
 	The height integration of the bremsstrahlung emissivity is:
 	\begin{equation}
 	F_{\nu}=2.4 \times 10^{10} T^{-\frac{1}{2}}\rho^{2}\exp(\frac{-h\nu}{KT}) H G_b (erg s^{-1} cm^{-2}Hz^{-1}) 
 	\end{equation}
 	and 
 	The bolometric bremsstrahlung flux is:
 	\begin{equation}
 	F_{bol}=5 \times 10^{20} T^{\frac{1}{2}}\rho^{2} H G_b(erg s^{-1} cm^{-2})
 	\end{equation} 
 	The bremsstrahlung luminosity of a CADF is given by:
 	\begin{equation}
 	L_{\nu}=2 \int F_\nu 2\pi r dr (erg s^{-1})
 	\end{equation}
 	by using the Schawrzchild units, the bremsstrahlung luminosity will be:
 	\begin{equation}
 	L_{\nu}=4\pi R^{3}_{s} \int_{1}^{R_{out}} F_\nu R^{2} dR
 	\end{equation}
 	As we mentioned in section 1, convective motions can transport energy from small to large radii and we can write the convective luminosity as $L_C\equiv \epsilon_c \dot{M} c^{2}$, where $\epsilon_c\cong 10^{-2}-10^{-3}$ is the convective efficiency. A fraction $\eta_{c}$ of this energy can be radiated at large radii in the CDAFs and this radiation emitted as thermal bremsstrahlung emission (Ball et al. 2001):
 	\begin{equation}
 	L_c=\eta_c \epsilon_c \dot{M} c^{2} 
 	 	\end{equation}
 	In this study we assume  $\eta_c=1$ which correspond to the all convected energy radiated away. The bolometric bremsstrahlung luminosity of a CDAFs is written as:
 	\begin{equation}
 	L_c=4\pi R^{3}_{s} \int_{1}^{R_{out}} F_{bol} R^{2} dR
 	\end{equation}
 	by equating this relation to $L_c=\eta_c \epsilon_c \dot{M} c^{2}$ and after some calculations we obtain :
 	\begin{equation}
 	\rho_0=0.37 (\frac{\eta_c \epsilon_c }{10^{-2}})\frac{10^{-3}}{m R^{2s-\frac{1}{2}}_{out}}\frac{(2s-\frac{1}{2})c_4}{c^{2}_0\sqrt{c_3}}
 	\end{equation}

Therefor, by substituting the above expressions for the density $\rho_0$ and temperature of the gas in equation 56, we find the bremsstrahlung spectrum from a CDAFs as follows:
\begin{displaymath}
\nu L_\nu=0.4 \times 10^{19}m  \frac{(\eta_c \epsilon_c)^{2}}{R^{4s-1}_{out}}\frac{c_4}{c^{2}_0 c^{\frac{3}{2}}_3}(2s-\frac{1}{2})^{2}
\end{displaymath}
\begin{equation}
\int_{1}^{R_{out}} \nu R^{2s-\frac{1}{2}} \exp{(-\frac{1.7\times 10^{-23}\nu R}{c_3})} dR
\end{equation}
We assume that the accretion flow extends from an outer radius $R_{out}$ down to an inner radius $R_{in}=1$. The bremsstrahlung emission can arises from all radii in the flow in contrast to synchrotron emission and modified Comptonization processes. As we have introduced above $\nu L_{\nu} \propto \nu \exp(\frac{-h\nu}{KT})$, by differentiating respect to $\nu$ from this expression, we can obtain the peak of the bremsstrahlung spectrum occurs at:
\begin{equation}
h \nu_{peak}=K T=K \frac{T_0 c_3}{R}
\end{equation}

As we can see in this equation, the peak of the spectrum is located in the X-ray frequency band.

\input{epsf}\epsfxsize=3.6in \epsfysize=2.3in\begin{figure}\centerline{\epsffile{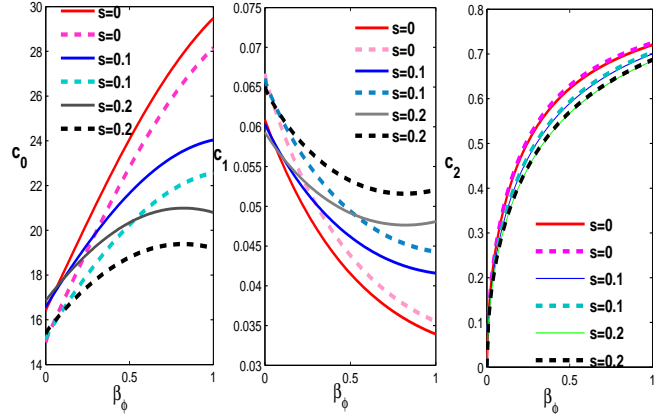}}
\caption{ Numerical coefficient $c_i$ as function of toroidal magnetic field
parameter $\beta_{\varphi}$ for several values of $s$ (the amount of the
wind/outflow). Solid lines correspond to model with $P_m=\infty$ (no magnetic diffusion) and dashed lines correspond to $P_m=0.5$.                        For all panels we use
$f=1$, $\beta_{r,z}=0.4$, $\alpha=0.5$, $\zeta=1.$, $\gamma=1.01$, $g=-\frac{1}{3}$. }
 \label{fig1}
\end{figure}

\input{epsf}\epsfxsize=3.6in \epsfysize=2.2in\begin{figure}\centerline{\epsffile{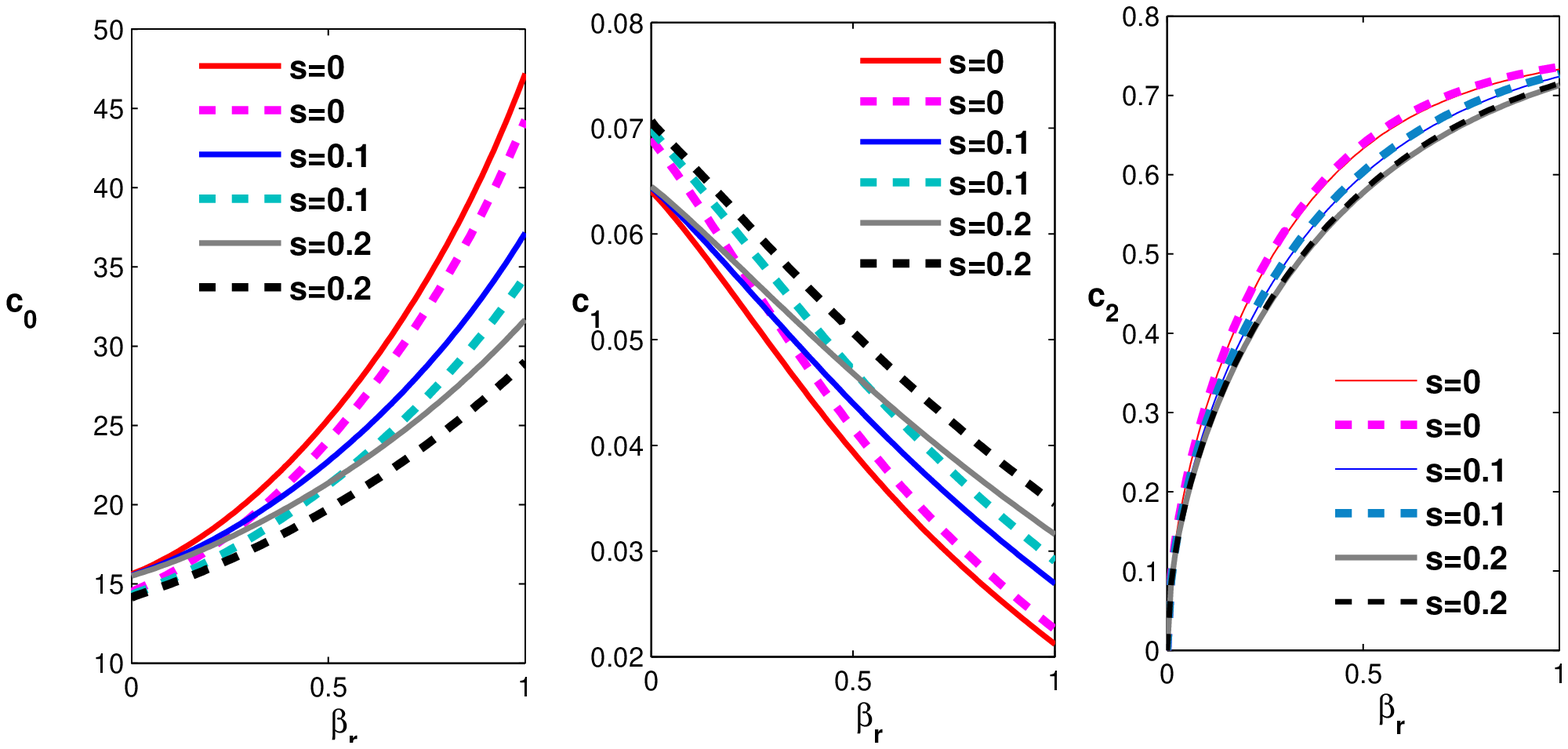}}
\caption{ Numerical coefficient $c_i$ as function of radial magnetic field
parameter $\beta_{r}$ for several values of $s$ (the amount of the
wind/outflow). Solid lines correspond to model with $P_m=\infty$ (no magnetic diffusion) and dashed lines correspond to $P_m=0.5$.                        For all panels we use
$f=1$, $\beta_{\varphi,z}=0.4$, $\alpha=0.5$, $\zeta=1.$, $\gamma=1.01$, $g=-\frac{1}{3}$. }
\label{fig2}
\end{figure}

\input{epsf}\epsfxsize=3.6in \epsfysize=2.2in\begin{figure}\centerline{\epsffile{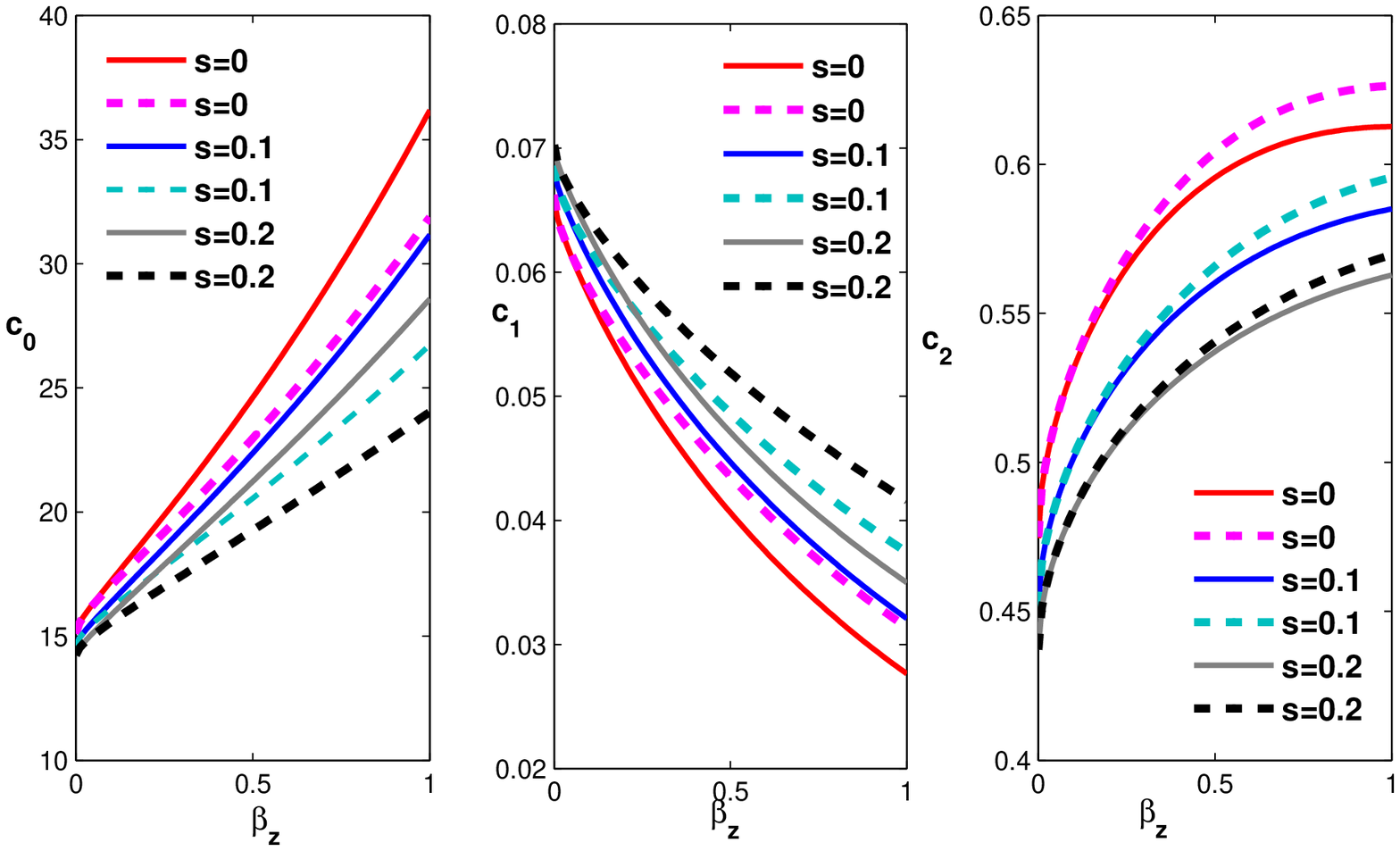}}
\caption{ Numerical coefficient $c_i$ as function of z-component of magnetic field
parameter $\beta_{z}$ for several values of $s$ (the amount of the
wind/outflow). Solid lines correspond to model with $P_m=\infty$ (no magnetic diffusion) and dashed lines correspond to $P_m=0.5$.                       For all panels we use
$f=1$, $\beta_{\varphi,r}=0.4$, $\alpha=0.5$, $\zeta=1.$, $\gamma=1.01$, $g=-\frac{1}{3}$. }
\label{fig3}
\end{figure}

\input{epsf}\epsfxsize=3.6in \epsfysize=2.2in\begin{figure}\centerline{\epsffile{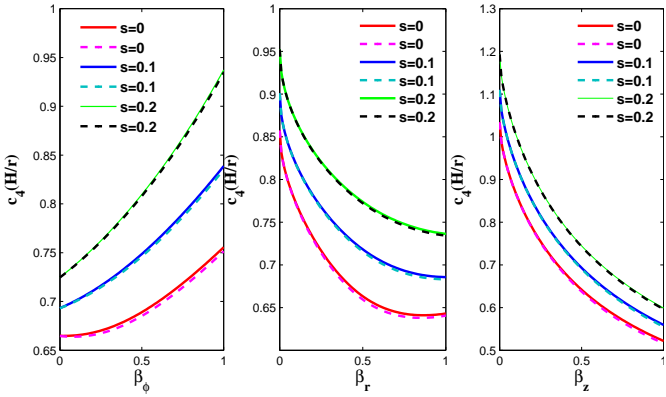}}
\caption{ The ratio of vertical thickness to radius $\frac{H}{r}$ as a function of toroidal magnetic field ($\beta_{\varphi}$, left panel, $\beta_{r,z}=0.4$), radial magnetic field ($\beta_{r}$, middle panel, $\beta_{\varphi,z}=0.4$) and z-component of magnetic field ($\beta_{z}$, right panel, $\beta_{\varphi,r}=0.4$ )
 for several values of $s$ (the amount of the
wind/outflow). Solid lines correspond to model with $P_m=\infty$ (no magnetic diffusion) and dashed lines correspond to $P_m=0.5$.                        For all panels we use
$f=1$, $\alpha=0.5$, $\zeta=1.$, $\gamma=1.01$, $g=-\frac{1}{3}$. }
\label{fig4}
\end{figure}

\input{epsf}\epsfxsize=3.6in \epsfysize=2.2in\begin{figure}\centerline{\epsffile{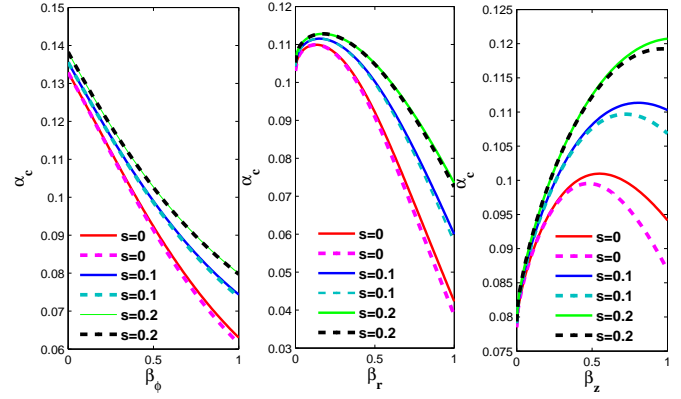}}
\caption{ The convection parameter $\alpha_c$ as a function of toroidal magnetic field ($\beta_{\varphi}$, left panel, $\beta_{r,z}=0.4$), radial magnetic field ($\beta_{r}$, middle panel, $\beta_{\varphi,z}=0.4$) and z-component of magnetic field ($\beta_{z}$, right panel, $\beta_{\varphi,r}=0.4$ )
 for several values of $s$ (the amount of the
wind/outflow). Solid lines correspond to model with $P_m=\infty$ (no magnetic diffusion) and dashed lines correspond to $P_m=0.5$.                        For all panels we use
$f=1$, $\alpha=0.5$, $\zeta=1.$, $\gamma=1.01$, $g=-\frac{1}{3}$. }
\label{fig5}
\end{figure}


\input{epsf}\epsfxsize=3.6in \epsfysize=2.8in\begin{figure}\centerline{\epsffile{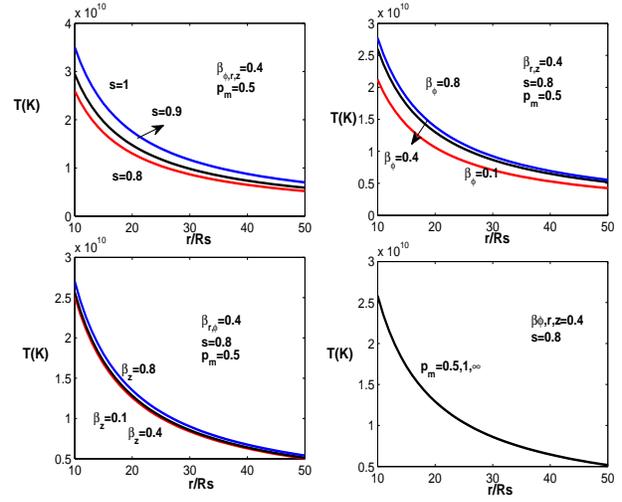}}
\caption{ This plot shows The surface temperature of the disc
$T(k)$ as function of dimensionless radius ($\frac{r}{R_\mathrm{s}}$)
for several values of (top left panel: wind parameter $(s)$, top right
panel: toroidal magnetic field parameter ($\beta_\varphi$), bottom left panel: z-component of magnetic
field parameter ($\beta_z$) and bottom right panel: magnetic Prandtl number ($P_m$)). For all panels we use
$f=1$, $\alpha=0.5$, $\zeta=1.$, $\gamma=1.01$,  $g=-\frac{1}{3}$. }
\label{fig6}
\end{figure}

\input{epsf}\epsfxsize=3.6in \epsfysize=2.8in\begin{figure}\centerline{\epsffile{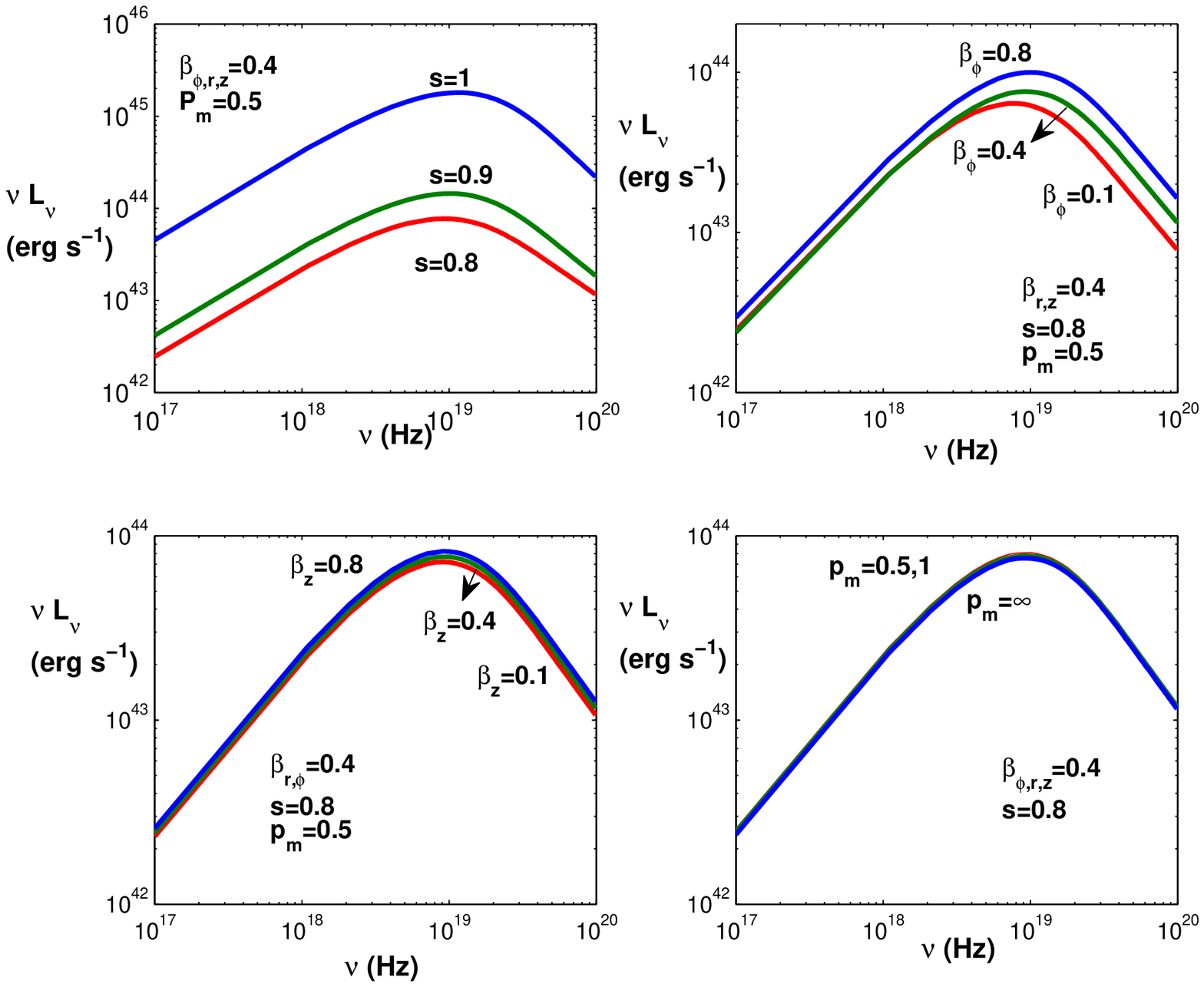}}
\caption{ Model CDAF spectra of a $10^{8} M_{\odot}$ black hole
for several values of top left panel: wind parameter $(s)$, top right
panel: toroidal magnetic field parameter ($\beta_\varphi$)($\beta_{r,z}=0.4$), bottom left panel: z-component of magnetic
field parameter ($\beta_z$)($\beta_{r,\varphi}=0.4$) and bottom right panel: magnetic Prandtl number $P_m$. For all panels we use
$f=1$, $\alpha=0.5$, $\zeta=1.$, $\gamma=1.01$,  $g=-\frac{1}{3}$. }
\label{fig7}
\end{figure}

\section{Results}

In this section, we will investigate numerically the role of magnetic field parameters, $\beta_{r, \phi, z}, P_{m}$ and also parameters, $s, l$, on the dynamical and observational appearance of CDAF in presence of all components of magnetic field. In this study constant values for some parameters like $\alpha=0.5$, $f=1$, $\gamma=1.01$, $g=-\frac{1}{3}$, $\zeta=1$, $\eta_c=1$ and $\epsilon_c=0.0045$ have been adopted.\\
 Figures 1,2,3 show the self similar coefficient $c_0$ (the surface density), $c_1$ ( the radial velocity) and $c_2$ (the rotational velocity) as a functions of the three components of magnetic field for different values of wind parameter $s (=0, 0.1, 0.2)$. By adding the all components of magnetic field $(\beta_{r,\varphi,z})$ which indicate the role of magnetic field on the dynamics of accretion disc, we see that the surface density and rotational velocity of the disc gradually increase, although the radial speed decreases. This results are qualitatively consistent with results presented by AM12 and FO12. In these figures we also studied the effect of parameter, $s$ which measures the strength of wind, on physical coefficients. Generally, when the exponent $s$ increases, the flow will rotate slower than that without wind. Also radial and rotational velocities in Figure 1, 2 and 3 represent significant deviations from non-wind solutions. Consequently, CDAFs in the presence of wind rotate more slowly than those without winds and wind leads to enhance accretion velocities. The strong wind causes the strong accretion velocity and reduction of surface density. For the larger values of $\beta_{r,\varphi}$, a small change in $c_2$ is observed. \\
 
 Our focus is the investigate the effects of magnetic resistivity (or the Prandtl number $P_m$) and outflow on the structure of the disc. Here, the inverse of Prandtl number specifies the resistivity of the fluid as we stated in section 2. Our solutions switch back to the solution in AM12, in which the magnetic resistivity is not considered (or $P_m=\infty$). Figure 1,2,3,4 and 5 show a comparison of our model and AM12. As is clear in these Figure 1, the influence of magnetic diffusivity on the surface density and accretion velocity is more evident for deferent values of $\beta_{\phi}$. The effects of magnetic diffusion and outflow are almost the same on the structure of our model. These properties confirm the results of Faghei \& Mollatayefeh (2012) and FO12. 
 
 The vertical thickness profiles are presented as a function of $\beta_{r,\varphi,z}$  for various values of exponent of $s$ in Figure 4. They show that the disc vertical thickness increases with increasing $\beta_{\phi}$, toroidal component,  but decreases by increasing the other components. This figure demonstrates that the disc thickness increases by increasing  s. Equation 39, is clearly shows that the vertical thickness posses a complicated dependencies with our input parameters. This figure reveal that the magnetic diffusion has not a obvious effect on the thickness of disc.

In figure 5. we have plotted the convection parameter $\alpha_c $ versus the components of magnetic field for several values of s. The results are compatible with AM12. As it is seen, the convective parameter decrease, if the radial and toroidal magnetic field parameter become stronger
although by adding z-component of magnetic field the convective parameter $\alpha_c$ increases. Larger values of wind parameter, $s$, cause the strong convection in the disc. It is obviously seen, the effect magnetic resistivity on the convection parameter is more important for the large vertical magnetic field. The results of wind and resistivity on the convection is the same.

In figure 6. the surface temperature ($T_{eff}$ ) is plotted as a function of the dimensionless radius ($\frac{r}{R_s}$).
It is obvious that the surface temperature is monotonically decreasing with  $\frac{r}{R_s}$. We see that the surface temperature increases by adding
wind parameter $s$, top left In Figure 6. In the top right panel and bottom left panel we
show that surface temperature increases by increasing toroidal and vertical
magnetic field parameters. But the effect of $\beta_{\varphi}$ is more evident. And finally in the bottom right panel
we can see the values of magnetic resistivity don't affect the surface temperature. So the disc is hotter with strong wind or even stronger magnetic field.
 
The radiation spectrum of the hot CDAF around a $10^{8} M_{\odot}$ black hole is represented in Figure 7 for several values of wind, radial and vertical magnetic field  and resistivity parameters. The corresponding values of $\dot{m}(=\frac{\dot{M}}{\dot{M}_{Edd}})$ and $R_{out}$ are $10^{-3.9}$ and $10^{3}$ respectively.
Luminosity depends explicitly on some of our input parameters like $s$, $R_{out}$, $m$, $c_4$ ( see equation 60) and implicitly on other parameters.
 The bremsstrahlung X-ray emission has been adopted since it is dominant mechanism in the outer parts of CDAFs.

As can be seen in figure 7, the maximum of $\nu L_{\nu}$ is highly depends on the given values of wind parameter. Whenever bremsstrahlung emission dominants, the X-ray emission depends directly on $\dot{m}_{out}(=\frac{\dot{M}_{out}}{\dot{M}_{Edd}})$ (see section 3.1 in Quataert \& Narayan 1999 and see equation 57 in our paper). As the wind becomes stronger, the inner mass accretion rate $\dot{m}_{in}(=\frac{\dot{M}_{in}}{\dot{M}_{Edd}} = \dot{m}_{out} R_{out}^{-s})$ becomes smaller than $\dot{m}_{out}$, so the importance of bermesstrahlung emission increases. So by adding the wind parameter $(s)$ the maximum of bermesstrahlung luminosity increases. Quataert \& Narayan (1999) stated the bremsstrahlung emission produces a peak that extends from a few to a few hundred $kev$.  Also we have shown that the peak of power spectrum is located approximately at $ h\nu \approx 41 keV $, which is independent of the strength of winds in the system. This result is in full agreement with perviously study by Quataert \& Narayan (1999).

Also, we are plotted the radiation spectrum of CDAFs for different values of all components of magnetic field. The effect of toroidal magnetic field on the radiation spectrum is similar to wind parameter. While the effect of $\beta_{z}$ and $p_m$ on the power spectrum are almost negligible.

\section{Summary and Conclusion}
The CDAFs model  consistently explain radiation inefficient accretion flow (RIAF) in the presence of convection. Many authors studied the importance of convection in RIAFs (CDAFs model) by means of numerical simulation or even analytically using the mixing length theory. Also observational results and MHD simulation confirm the importance of outflow and magnetic diffusion in CDAFs. Our primary focus in this research is to develop of AM12 solutions by considering magnetic resistivity in the main MHD equations. Folowing FO12, AM12 and Zhang \& Dai 2008 we considered power-law function for mass inflow rate and solved the inflow-outflow equations by using self-similar approach in CDAFs regime. Some approximation have been done in order to simplify the main equations. We ignore the relativistic effects, self-gravity of the discs. For viscosity $\alpha$-prescription has been adopted. Our results reduce to AM12 solutions when the effect of magnetic resistivity is neglected. 

Consequently our results represent that by increasing all components of magnetic field the surface density and rotational velocity increase although the radial velocity decreases. Also existence of the wind will lead to a significant reduction of surface density  as well as rotational velocity and increasing radial velocity. Increasing $\beta_{\phi}$ will increase vertical thickness while it decreases by increasing $\beta_{z, r}$. Additionally the radial velocity and vertical thickness is will increase when outflow becomes important while the surface density and rotational velocity will decrease. Our results shown that the radial and rotational velocity will increase if the magnitude of resistivity increases, while rotational velocity decreases. The influence of magnetic diffusivity on the surface density and accretion infall velocity
is more evident in the bigger toroidal magnetic field. These results are generally constant with results presented by AM12, FO12.

We also calculated the continuum spectrum emitted from the discs with assuming bremsstrahlung mechanism. Convection motion in CDAF transports a large amount of energy stored in small radii to large radii. Igumenshchev \& Abramowicz (2000) suggested that some or perhaps main part of this energy budget might be radiated a way to the outer regions as a thermal Bremsstrahlung emission. So we used this model in our calculations. Consequently our solutions show that the bolometric luminosity increases as wind/outflow becomes stronger. The maximum of $\nu L_{\nu}$ is strongly changed by different values of wind parameter. On the other hand, increasing $\beta_{\phi}$ make the bolometric luminosity of the disc increase gradually. The other components of magnetic field and magnetic resistive have not a considerably effect on power spectra.

The main feature of the self-similar solution is that it is purely analytic and provides a transparent way of understanding the key properties of an CDAFs. However, the self-similar solution is not valid near the inner or outer boundaries. Consequently for calculating the radiation spectrum which mostly comes from inner regions (where the self-similar solution is invalid), we requires a global solution  obtained by solving directly
the differential equations of the problem. on the other hand in this paper we only considered the bremsstrahlung emission; while in the inner region of the disks, where the radiation mainly comes from there, the synchrotron radiation and its Comptonization is much more important than bremsstrahlung emission. Thus it worth to investigations the effects of these mechanism of radiation with global solutions.

Although that we have made some simplification and some assumption in order to solve equations analytically, our solutions show explicitly that outflow, large scale magnetic field and its corresponding resistivity can really change dynamical and observational appearance of CDAF. It means in any realistic model these parameters should take into account. This kind of self-similar solution could greatly facilitate testing and interpretations of the numerical simulations and observational evidences.
 
 S. Abbassi acknowledges support from the International Center for Theoretical Physics (ICTP) for a visit
through the regular associateship scheme.

\end{document}